\documentstyle[12pt]{article}
\def\mt{ m_t  }  \def\MW{ m_W }   \def\GeV{  GeV }  
\def\Cite{ref.\ \cite}

\begin{document}
\begin{center}
{\LARGE\bf QCD Corrections for the} \\
\vspace{3mm}
{\LARGE\bf $K^0 - \bar K^0$ and $B^0 - \bar B^0$ System} \\
\vspace{15mm}
{\Large A. Datta, E.A. Paschos, J.-M. Schwarz} \\
\vspace{3mm}
Institut fuer Physik, Universitaet Dortmund \\
D-44221 Dortmund, Germany \\
\vspace{5mm}
and \\
\vspace{5mm}
{\Large M.N. Sinha Roy} \\
\vspace{3mm}
Dept. of Physics, Presidency College \\
Calcutta 700 073, India
\end{center}
\vspace{22mm}

\begin{abstract}
The QCD corrections for the box diagrams are revisited for the case
of a heavy top quark with $\mt = 174 \GeV$.  We resolve first a
longstanding discrepancy between two methods of calculation by
showing that they give the same results when the threshold factors
are treated correctly.  Using this observation we refine our earlier
results and derive formulae valid for the $K$- and $B$-meson systems.
Our formulae are given in terms of integrals to be evaluated numerically,
as well as approximate analytical formulae. These calculations include
the evolution above $\MW$ which has been neglected by other authors.
\end{abstract}

\newpage
\section{Introduction}

\label{sec:intro}

Perturbative quantum chromodynamic (QCD) corrections to electroweak
processes,
dominated by short distance physics, were calculated in the leading
logarithmic approximation (LLA) long time ago \cite{R1}.
The pioneering calculations studied the enhancement of the $\Delta I
= 1/2$-amplitude in $\Delta S = 1$ processes and the origin of direct
CP violation.

Subsequently, QCD corrections in the LLA were computed for
$\Delta S = 2$ processes \cite{R2}-\cite{R4} assuming that all the
quarks in the loops are much lighter than the W bosons ($m_q \ll \MW$
for all $q$).  In  course of time, the gradual strengthening of
the lower limit
on the mass $\mt$ of the t quark lead to the realization that $\mt$
may indeed be comparable to $\MW$ or even larger.  This motivated several
papers which generalize the calculations for a heavy top quark
\cite{R5}-\cite{R9}.
Some of these articles \cite{R6}-\cite{R9}
simpified the calculation by
neglecting the evolution of the Wilson coefficients in the interval
$\MW \to \mt$.  This approximation was not made in \cite{R5}.
However, since $\mt$ was unknown, the numerical significance of this
approximation could not be assessed properly.\bigskip

In the meanwhile, the production of the top quark has been reported
by two Fermilab groups.  The CDF collaboration \cite{R10} finds a
top quark with a mass
$\mt = 174 \pm 10 \pm {13\atop12} \GeV $
and the D0 collaboration \cite{R11} with a mass
$\mt = 180 \pm 12 \GeV $ .
These values are in good agreement with the precision measurements
at LEP-I where it was established \cite{R12,R13} (albeit indirectly) that
$\mt = 169 \pm {16\atop18} \pm {17\atop20} \GeV $.
All this forces us to accept that  the mass difference between the top
quark  and   the  W-boson is  indeed   significant.   It  is therefore
worthwhile  to revisit the   calculations of \cite{R5} and compare them
with other results  \cite{R6}-\cite{R8} in order to estimate the  numerical
significance    of  the evolution beyond   $\MW$.    We find that this
evolution  is   numerically  significant for  $\mt  \approx  174 \GeV$
although it does not change the result drastically.

The first isssue concerns the  method of calculation to be adopted
in this article.  There are two methods : the first one is diagramatic
where one loop  QCD corrections are computed on top of the weak box diagram.
The lowest order corrections thus obtained, are summed
up using the renormalisation group (RG). The resulting Wilson coefficients
are dependent on the momentum of the electroweak loop ($p^2$) which
are then integrated over $p^2$ \cite{R2,R4,R5}.  The
second method first introduced in \cite{R3} computes the QCD
corrections in a series of effective field theories obtained by successively
integrating out the heavy degrees of freedom \cite{R3,R6,R7,R8}. The two
methods have a long-standing and unsettled issue which was raised in
\Cite{R3}. To be specific,even in  case of $\mt \ll \MW$,
the results of \cite{R3} (which heavily uses the rather
involved techniques of operator mixing) and \Cite{R4} did not
agree.  Indeed it was noted in \Cite{R3} that although two of the three
QCD correction factors($\eta_1, \eta_2$) computed by them agree with
\Cite{R4} after appropriate simplifications, the third ($\eta_3$)
did not (see footnote 14 of
\Cite{R3}).  To the best of our knowledge this discrepancy has not
been clarified in the subsequent literature.  Since the disagreement
was precisely in a term where operator mixings played an important
r\^ole in \Cite{R3}, a critical reader may also question the procedure
of \Cite{R4},
which did not
explicitly use operator mixing.
This critique would also apply to
\Cite{R5} where the techniques of Novikov, Shifman, Vainshtein, Zakharov
\cite{R2} and Visotski{\u \i} \cite{R4} (NSVZV) was generalized for
$\mt \gg \MW$.

In this paper we explicitly demonstrate that the apparent disagreement
between \Cite{R3} and \cite{R4} is a consequence of certain
simplifying assumptions in \Cite{R4}.  More specifically in \Cite{R4}
certain Wilson coefficients were evolved over  large momentum
intervals  which involve several thresholds.
Some remarks were made regarding the ambiguities in
determining the number of active quark flavours ($n_f$) in these
intervals, but for the purpose of actual computation, $n_f$ was kept
fixed.  We find that by changing $n_f$ at each new quark threshold and
taking the matching conditions at each threshold \cite{R4} into account a more
rigorous formula can be derived from the procedure of \Cite{R4}.  This
leads to an agreement --- not only numerically but also algebraically
--- with \Cite{R3}.

Although the case $\mt \ll \MW $ is now of academic interest only , we
consider the above demonstration important because it illustrates that both
methods of calculation (GW and NSVZV and its generalization \cite{R5})
are equivalent.  The apparent disagreement comes from the approximate
treatment of the threshold factors as discussed above.  In the present
article, we remove these approximations and sharpen the formulae of \Cite{R5}.
In section 3 we present explicit and improved formulae which are in
one-to-one correspondence to our previous work \cite{R5}.  The formulae
by themselves answer several questions raised about our earlier work
(see in particular \cite{R8}).

The main new result of  \Cite{R5} was to demonstrate that a particular
Wilson coefficient
computed in \cite{R2,R4} does not contribute if $m_t >> m_W$.
The remaining Wilson coefficients
were calculated. The $p^2$ dependent Wilson coefficients as discussed
above were integrated numerically over the entire permissible range of
$p^2$. In this paper we incorporate the  threshold factors discussed
earlier, into these integrations and obtain very accurate results.

However, from the final results thus obtained the numerical significance
of the evolution of $\alpha_s( p^2)$ over a specific interval
$(\MW \longrightarrow \mt )$ cannot be isolated automatically.
 This can be done by separating the regions of the numerical integration
and setting $\alpha_s(p^2) = \alpha_s(\MW^2)$ for $p^2 > \MW^2$.
 Alternatively, in order to
separate the evolution of this specific interval, we introduce a reasonable
approximation, which allows us to perform the $p^2$-integration
analytically.  This way we isolate the interval $(\MW \longrightarrow
\mt )$, where we can choose $\alpha_s(p^2)$ as either running or as
a constant with $\alpha_s(p^2) = \alpha_s(\MW^2)$ for $p^2 > \MW^2$
. We thus obtain for the first time a set of analytical formulae
which explicitly exhibits the effects of the evolution above $m_W$
and readily reduces to the results of \cite{R6,R7,R8} once these
evolutions are neglected. However, as we summarize below in various tables,
 the evolution above $\MW$,though not completely negligible, does not
change the results drastically. This is partly due to the fact that
various corrections cancel each other. Finally, the formalism of
this article is very transparent and general so that it can be easily
extended to four generations, where higher scales become relevant.
In such cases the evolution above $m_W$ can in principle  be crucially
important. It has been shown that  models with naturally
 heavy Majorana \cite{R14} or Dirac \cite{R15} neutrinos
 belonging to the fourth generation can be constructed.
 Such models are ,therefore, neither unnatural nor in conflict with
 the neutrino counting at LEP\cite{R12,R13}.

In \Cite{R8} next to leading order calculations for one of the QCD
factors, $\eta _2$ in the notation of \cite{R3}, have been performed.
These attempts are important since they have the potential of settling
important theoretical issues, like the independence of the final result
from the scale at which the initial conditions for the Wilson
coefficients have been imposed or from the definitions of the heavy
top masses, etc.  Such refinements however can attain their full
potential only if the corrections in the LLA are known as accurately
as possible.  In this paper we find that in some cases
the evolution above $\MW$ is numerically of the same order as the
nonleading corrections and reqires, therefore, a closer scrutiny
(see below).  This explains, in part, the
fact that the authors in \cite{R8} obtained for $\eta _2$ the same result.
For the case $\mt>\MW$ there is for $\eta _3$ only the calculation
in \cite{R5} and the present one.

The paper is organized as follows.  In section 2 we briefly review
the well known basic formulae for $\Delta F = 2$ processes relevant
for $K^0 - \bar{K}^0$ and $B^0 - \bar{B}^0$ mixings in order to set up
the notation and make the article self-contained.  We then review
the computation of the QCD corrections in the hypothetical case
$\mt \ll \MW$ following the NSVZV method and demonstrate the equivalence
with GW.  This is reached by treating the threshold factors and the
matching conditions carefully.  Using the experience gained in section
2, we refine the results of \Cite{R5} and present them as specific
integrals.  This is done explicitly for each integral of the effective
weak Hamiltonian and summarized in section 3.  In addition, section 3
includes analytic formulae obtained after integrating the terms, in a
certain approximation, in order to separate the evolution above $\MW$
and compare it with other works \cite{R6} -- \cite{R8}.  The formulae
look long and complicated, but are obtained by a straight-forward
treatment of thresholds.  In section 4 we evaluate the results numerically
arriving at final values for the QCD factors and compare them with
those in other articles.

\section{QCD Corrections in the NSVZV approach}
\label{secII}

In this section we consider the hypothetical case $ \MW \gg m_q$ (for
all $q$) in order to connect with the early work on this subject and
in particular to demonstrate the equivalence between the works of
Gilman and Wise \cite{R3} and Visotski{\u\i} \cite{R4}.

In this limit  ($\MW \gg m_q $) the  effective Hamiltonian without QCD
corrections  is obtained from  the box diagram (Fig.\ 1) with
two W boson exchanges and attains the form \cite{R16}:
\begin{eqnarray}
  \label{eq:1}
  H_{eff}  & =&     \frac{G_F^2  \MW^2}{16\pi^2}  (\bar    d  \gamma_\mu
  (1-\gamma_5)s)^2 \sum_{i,j=c,t} \lambda_i \lambda_j [ I(m_i^2,m_j^2)
  - I(m_i^2,0) - I(0,m_j^2) + I(0,0) ] \nonumber\\
      &\equiv& \frac{G_F^2  \MW^2}{16\pi^2}  (\bar    d  \gamma_\mu
  (1-\gamma_5)s)^2 \sum_{i,j=c,t} \lambda_i \lambda_jH_{eff}(i,j)
\end{eqnarray}
where  the GIM subtraction  is explicitly introduced with $m_u \approx
0$.  We follow  the standard notation  $\lambda_i  = V_{id}^* V_{is}$,
where  the $V$'s  are the  elements  of the Cabibbo-Kobayashi-Maskawa
 (CKM)
matrix.   The  dimensionless     integrals $I(m_i^2,m_j^2)$   are  over
Euclidean loop momentum,

\begin{equation}
  \label{eq:2}
  I(m_i^2, m_j^2)  =
      \int  \frac{dp^2 p^4 \MW^2}
                 {(p^2+m_i^2)(p^2+m_j^2)(p^2+\MW^2)^2}
\end{equation}

These integrals are convergent so that the dominant contributions come
from $p^2 \approx m_i^2,m_j^2 $ as well as $p^2 \approx \MW^2$.  Due
to the GIM subtraction in Eq. (\ref{eq:1}) the contribution form
the region $p^2 \approx \MW^2$ cancels out. Hence it is justified to
replace the W propagators by $\MW^{-2}$.  Now $H_{eff}$ is implicitly
the time-ordered product of two $\Delta S = 1$, local four fermion
operators.  This of course can be seen most transparently in
coordinate space (see Eqs. 5a--5c of \Cite{R3}). These operators
can be written as linear combinations of the well known colour
symmetric and antisymmetric operators $O^\pm$.  The QCD corrections
for these multiplicatively renormalisable operators are well known
\cite{R1}.  Finally the QCD corrected $\Delta S = 1$ operators
(neglecting mixings with penguins) are joined again by Wick's rule.
In momentum space this corresponds to the replacement of the W boson
propagators in Eq. (\ref{eq:2}) by $\MW^{-4}$ and multiplying
each integral
of Eq.( \ref{eq:1}) by a factor \cite{R2,R4}
\begin{equation}
   \label{eq:3}
  Q_1(\MW^2,p^2,n_1)                       =                   \frac12
  (\frac{\alpha_{n_1}(\MW^2)}{\alpha_{n_1}(p^2)})^{\frac{2a_-}{b(n_1)}}
  -
  (\frac{\alpha_{n_1}(\MW^2)}{\alpha_{n_1}(p^2)})^{\frac{a_++a_-}{b(n_1)}}
  +                                                            \frac32
  (\frac{\alpha_{n_1}(\MW^2)}{\alpha_{n_1}(p^2)})^{\frac{2a_+}{b(n_1)}}
\end{equation}
where $a_+ = 2$, $a_- = -4$ are the anomalous dimensions of $O^\pm$ in
units of  $\alpha_{n_1}/(2\pi)$, $\alpha_{n_1}$ is the  strong  coupling
constant and  $b(n_1) = 11  -  \frac23 n_1$ depends  on  the number of
massless quarks  in the interval $\MW  \to  p^2 $.  We  emphasize that
since $p^2$  is variable   $n_1$  changes as $p^2  $   crosses various
quark thresholds.

The two momentum scales in Eq. (\ref{eq:3}) are worth noting.  It
was emphasized in \Cite{R2} that for Green functions involving loops,
$p^2$ provides an important scale and the emergence of momentum ($p^2
$) dependent Wilson coefficients was explicitly demonstrated ( see in
particular the third article of \Cite{R2}).
Moreover, as pointed out in \Cite{R2,R4}, and directly shown in
\Cite{R5}, the $O(\alpha)$ terms in the expansion of Eq.
(\ref{eq:3}) can be directly identified with the diagrams of figure 1
(see Eqs. 20--21b of \Cite{R5}).

Once $p^2 $ becomes larger than the masses and momenta on the external
legs,   the effective  four   fermion  operator  $(\bar  d  \gamma_\mu
(1-\gamma_5)s)^2$ is generated.  This property  is already implicit in
Eq.( \ref{eq:1}),     which  is derived   in  the  approximation of
neglecting the  external momenta  and  it also holds in  the realistic
case $\mt > \MW $.  The evolution of  the four fermion operator $(\bar
d \gamma_\mu (1-\gamma_5)s)^2$, which is colour symmetric and has the
same anomalous dimension as $O^+$, from $p^2$ down to $\mu^2$ (a scale
where perturbative QCD presumably breaks down) yields a factor
\begin{equation}
   \label{eq:4}
  Q_2(p^2,\mu^2,n_2)  =
  (\frac{\alpha_{n_2}(p^2)}{\alpha_{n_2}(\mu^2)})^{\frac{2}{b(n_2)}}
\end{equation}
which multiplies  the integrands of Eq. (1).  The $O(\alpha)$ term
in the expansion of Eq. (\ref{eq:4}) can be explicitly checked from
fig.\ 2  as was  done in \Cite{R5}   (see Eq. (12)
of \cite{R5} and the
discussion following it). It is very  important to note
here that the number of active flavours ($n_2$) in Eq. (\ref{eq:4})
(which  is not  fixed as yet) can  be different  from  that in Eq. (3)
since the momentum intervals in the two cases are different.

Finally each  of  the  heavy quarks  in  the  internal lines  are  off
mass-shell during the $p^2$ integration and can be dressed with gluons
as  in figure  3. This corresponds   to replacing each  heavy
quark mass in Eq. (\ref{eq:1}) by \cite{R2,R4}
\begin{eqnarray}
  m_q^2(p^2) &=& m_q^2(m_q^2)
      (1 + b(n_3) \frac{\alpha_{n_3}(m_q^2)}{4\pi}
       \ln\frac{p^2}{m_q^2} ) ^ {-8/b(n_3)}
                                \nonumber\\
       \label{eq:5}
       &=&   m_q^2(m_q^2)
  (\frac{\alpha_{n_3}(p^2)}{\alpha_{n_3}(m_q^2)})^{8/b(n_3)}
\end{eqnarray}
which  is  the well known relation    between running quark  masses at
scales $p^2$  and $m_q^2$.
The right-hand side of  Eq. (\ref{eq:5}) was not shown
explicitly in \Cite{R2,R4,R5},  but was  evidently
defined correctly.   This   caused  some criticism in the
subsequent literature \cite{R8}, which is not justified.

Following our  above observations we have kept  the number of flavours
($n_3$  in Eq. (\ref{eq:5}))  arbitrary and  in principle different
from those in Eqs.  (\ref{eq:3},\ref{eq:4}).  It is rather  obvious
that this correction is  also independent of  the condition $\MW^2 \gg
\mt^2$ and continues to hold in the realistic case $\mt > \MW $.

The removal of the W  boson propagators and GIM subtraction leads
to a generic contribution to the effective Hamiltonian defined in Eq.( 1)
(in the absence
of QCD corrections)
\begin{equation}
  \label{eq:6}
  H_{eff}(i,j)  \approx     \frac1{\MW^2}     \int   \frac{m_i^2       m_j^2
    dp^2}{(p^2+m_i^2)(p^2+m_j^2)} .
\end{equation}

When two quarks on the internal lines of the box diagram are identical
($ i = j =  c $ or $  t $, henceforth referred  to  as the  cc or tt
graph) the integral  is dominated by  contributions from  $p^2 \approx
m_i^2$   \cite{R2,R4}.  On the other hand,    for the diagram with one
internal c quark and one  t quark (the ct graph),  the entire region $
m_c^2 \le p^2   \le \mt^2  $  contributes  as can  be  seen  from  the
following form \cite{R2,R4}
\begin{eqnarray}
  \label{eq:7}
  H_{eff}(c,t)        &\approx&     \frac{m_c^2  \mt^2} {\MW^2(\mt^2-m_c^2)}
  \int_{m_c^2}^{\mt^2}  \frac{ dp^2}{p^2  }  \nonumber\\
  &\approx& x_c \ln\frac{m_t^2}{m_c^2}
\end{eqnarray}
where $ x_q = m_q^2/\MW^2$.

Due to QCD corrections the above formula should be modified to
\begin{equation}
  \label{eq:8}
  H_{eff}(i,j) = \frac1{\MW^2}  \int \frac{dp^2 m_i^2(p^2) m_j^2(p^2)
    Q_1(\MW^2,p^2,n_1)      Q_2(p^2,\mu^2,n_2)}      {(p^2+m_i^2(p^2))
    (p^2+m_j^2(p^2))}
\end{equation}
where the corrections are defined in Eqs. (\ref{eq:3})--(\ref{eq:5}).
A very  accurate analytical result  can be  obtained by  the following
approximation.  It  is assumed that   $p^2$ dependent QCD  corrections
cannot drastically alter  the important regions of momentum integration
as  noted after  Eq. (\ref{eq:6}).  For the $i  = j$ case, this
corresponds       to       evaluating   the   corrections   (Eqs.
\ref{eq:3}--\ref{eq:5}) at $p^2 =  m_i^2(m_i^2)$ and one obtains from
Eq. (\ref{eq:8})

\begin{equation}
  \label{eq:9}
  H_{eff}(i,i) = x_i Q_1(\MW^2,m_i^2,n_1) Q_2(m_i^2,\mu^2,n_2)
\end{equation}

\noindent For notational  convenience we shall  denote  henceforth
$m_i(m_i^2) = m_i$.

In \Cite{R4}, $n_2$ in Eq. (\ref{eq:4}) was assumed to be 3 for
numerical calculations.  While this is justified for  the cc graph the
same cannot be said about the tt graph.  Similarly $n_1$ was chosen to
be 4 which is not rigorously justified either for the cc or for the
tt graph.  For  the cc graph, for  example, one can replace a  typical
term in $Q_1$ of Eq. (\ref{eq:9}) by
\begin{equation}
  \label{eq:10}
 (\frac{\alpha_{n_1}(\MW^2)}{\alpha_{n_1}(m_c^2)})^{a/b(n_1)} \longrightarrow
  (\frac{\alpha_{6}(\MW^2)}{\alpha_{6}(m_t^2)})^{a/b(6)}
  (\frac{\alpha_{5}(m_t^2)}{\alpha_{5}(m_b^2)})^{a/b(5)}
  (\frac{\alpha_{4}(m_b^2)}{\alpha_{4}(m_c^2)})^{a/b(4)}
\end{equation}
where the  changes in the number of   flavours at different thresholds
have been  explicitly   included.  It is further assumed that the matching
conditions as defined in \cite{R3} are also obeyed.
With  this   modification  Eq.
(\ref{eq:9}) reproduces the results of \Cite{R3}, which was already noted
in \Cite{R3}.   A similar agreement for the  tt  graph can be obtained
analogously by modifying the $Q_2$ term in Eq. (\ref{eq:9}).

The situation  is somewhat more complicated for   the ct graph.  Using
Eq. (\ref{eq:7}), we can  write the integral of Eq. (\ref{eq:8})
in the following approximate form
\begin{equation}
  \label{eq:11}
  H_{eff}^{QCD}     (c,t)     =    \frac1{\MW^2}  \int_{m_c^2}^{\mt^2}
  \frac{dp^2}{p^2} m_c^2(p^2) Q_1(m_W^2, p^2, n_1) Q_2(p^2, \mu^2, n_2)
\end{equation}

When we substitute in  Eqs. (\ref{eq:3})--(\ref{eq:5}) the number
of
flavours $n_{1,2,3} = 4,3,4$, respectively, and factorize
\begin{displaymath}
\frac{\alpha (p^2)}{\alpha(\mu ^2)} =
\frac{\alpha (p^2)}{\alpha(m^2_c)}
\frac{\alpha (m^2_c)}{\alpha(\mu ^2)}
\end{displaymath}
and
\begin{displaymath}
(\frac{\alpha(\MW^2)}{\alpha(p^2)}) =
(\frac{\alpha(\MW^2)}{\alpha(m_c^2)})
(\frac{\alpha(m_c^2)}{\alpha(p^2)})
\end{displaymath}

\noindent which roughly corresponds to the approximation that above $m_c$
the number of active quark  flavour  is fixed at 4,  the integration in
Eq. (\ref{eq:11}) can  be done easily  and $\eta_3$ of \Cite{R4} is
rederived. The resulting formula unfortunately disagrees with \Cite{R3}.
This is the disagreement already mentioned and will be resolved below.

In order to treat the number of  active flavours consistently, we note
that a typical term of Eq.( \ref{eq:11}) is of the following form:
\begin{equation}
  \label{eq:12}
  I_a            =        \int_{m_c^2}^{\mt^2}        \frac{dp^2}{p^2}
  (\frac{\alpha_{n_3}(p^2)}{\alpha_{n_3}(m_c^2)})    ^      {(8/b(n_3))}
  (\frac{\alpha_{n_2}(p^2)}{\alpha_{n_2}(\mu^2)})    ^      {(2/b(n_2))}
  (\frac{\alpha_{n_1}(p^2)}{\alpha_{n_1}(\MW^2)}) ^ {(a/b(n_1))}
\end{equation}

Values for $n_1$, $n_2$, and  $n_3$ can be assigned consistently  by
splitting the  integral into two pieces  $m_c^2 \le p^2   \le m_b^2  $
and $m_b^2 \le p^2   \le \mt^2  $ . Now each momentum interval can be
assigned  a well defined number of active flavours.  This yields
\begin{eqnarray}
  \label{eq:13}
  I_a   &=&
     (\frac{\alpha_{3}(m_c^2)}{\alpha_{3}(\mu^2)}) ^ {(2/b(3))}
     (\frac{\alpha_{6}(\mt^2)}{\alpha_{6}(\MW^2)}) ^ {(a/b(6))}
     \nonumber\\&& 
     \left\{
       (\frac{\alpha_{5}(m_b^2)}{\alpha_{5}(\mt^2)}) ^ {a/b(5)}
       (\frac{\alpha_{4}(m_c^2)}{\alpha_{4}(m_b^2)}) ^ {a/b(4)}
       \int_{m_c^2}^{m_b^2}  \frac{dp^2}{p^2}
         (\frac{\alpha_{4}(p^2)}{\alpha_{4}(m_c^2)}) ^ {(8+a+2)/b(4)}
     \right.\nonumber\\&&\left. 
         +
       (\frac{\alpha_{4}(m_b^2)}{\alpha_{4}(m_c^2)}) ^ {10/b(4)}
       (\frac{\alpha_{5}(m_b^2)}{\alpha_{5}(\mt^2)}) ^ {a/b(5)}
       \int_{m_b^2}^{\mt^2}
       \frac{dp^2}{p^2}
         (\frac{\alpha_{5}(p^2)}{\alpha_{5}(m_b^2)}) ^ {(10+a)/b(5)}
     \right\}
\end{eqnarray}
where a= -4, 2 or 8 and replacements similar to Eq.( 10) have been made.

Using the general formula
\begin{eqnarray}
  \label{eq:14}
  J(m_2^2,m_1^2,x)
  &=&
  \int_{m_1^2}^{m_2^2} \frac{dp^2}{p^2}
  \left(1+\frac{b(n)}{4\pi}\alpha_n(m_1^2)
    \ln\frac{p^2}{m_1^2}
  \right)^{-x}
  \nonumber\\
  &=&
  \frac{4\pi}{b(n)\alpha_n(m_2^2)(1-x)}
  (\frac{\alpha_{n}(m_2^2)}{\alpha_{n}(m_1^2)}) ^x
  \left(1 - (\frac{\alpha_{n}(m_1^2)}{\alpha_{n}(m_2^2)})^ {x-1}\right)
\end{eqnarray}
the integrals in Eq.(\ref{eq:13}) can be done analytically.

In  simplifying the   results,   the  matching conditions (e.~g.\    $
\alpha_4(m_b^2)    = \alpha_5(m_b^2) $  \cite{R3})  should   be used in
particular for the $\alpha_n(m_2^2)$ in the overall factor in Eq.
(\ref{eq:14}).  This leads to, e.g.
\begin{equation}
  \label{eq:15}
  \alpha_5(\mt^2) =
    (\frac{\alpha_{5}(\mt^2)}{\alpha_{5}(m_b^2)})
    (\frac{\alpha_{4}(m_b^2)}{\alpha_{4}(m_c^2)})
    \alpha_{4}(m_c^2)
\end{equation}

Using all of the above one finally obtains
\begin{eqnarray}
  \label{eq:16}
  I_a  &=&
    (\frac{\alpha_{3}(m_c^2)}{\alpha_{3}(\mu^2)})  ^ {2/b(3)}
    (\frac{\alpha_{6}(\mt^2)}{\alpha_{6}(\MW^2)})  ^ {a/b(6)}
    \frac{4\pi}{\alpha_4(m_c^2)(b(4)-10-a)}
    \nonumber\\&& 
    \left[
       \left(1 - \frac{b(4)-10-a}{b(5)-10-a}\right)
       (\frac{\alpha_{4}(m_b^2)}{\alpha_{4}(m_c^2)})  ^ {10/b(4)-1}
       (\frac{\alpha_{5}(\mt^2)}{\alpha_{5}(m_b^2)})  ^ {-a/b(5)}
    \right.\nonumber\\&&\left. 
    -
       (\frac{\alpha_{5}(\mt^2)}{\alpha_{5}(m_b^2)})  ^ {-a/b(5)}
       (\frac{\alpha_{4}(m_b^2)}{\alpha_{4}(m_c^2)})  ^ {-a/b(4)}
    \right.\nonumber\\&&\left. 
    +
       (\frac{b(4)-10-a}{b(5)-10-a})
       (\frac{\alpha_{4}(m_b^2)}{\alpha_{4}(m_c^2)})  ^ {10/b(4)-1}
       (\frac{\alpha_{5}(\mt^2)}{\alpha_{5}(m_b^2)})  ^ {10/b(5)-1}
    \right]
\end{eqnarray}

In this notation Eq. (\ref{eq:11}) can be rewritten as

\begin{equation}
  \label{eq:17}
  H_{eff} (c,t) = x_c
  \left[ \frac12 I_8 - I_2 + \frac32 I_{-4} \right]
\end{equation}

\noindent which  is  algebraically identical to   Eq.  (23) of \Cite{R3}.  In
summary we reiterate  that the effects  of the mixing of the operators
$O_7$, $O_8$  in the coordinate space  as studied in \Cite{R3}, can be
equivalently treated in this   momentum space calculation  by handling
the $p^2$ integration properly (notice the mixing of the anomalous
dimensions of various operators in the integrands of Eq. (13)).
Consequently, one can use the methods
of GW or  NSVZV and obtain the  same results, provided  that threshold
effects are handled  appropriately.   In the next section  we  include
these   effects in the  approach of  NSVZV  in order to refine the
results of \Cite{R5} for the case of $\mt \gg \MW$.
\section{QCD corrections in LLA for $\mt \gg \MW$}
\label{secIII}

We now present the QCD correction factors for the realistic case
$\mt \gg \MW $ which is a refined version of the calculations of \Cite{R5}
along the lines developed in section \ref{secII}. In this calculation
diagrams with unphysical Higgs scalar exchanges are also important.
In the following we denote by the subscripts WW, WH and HH the contributions
of diagrams with two W boson internal lines, one W and one Higgs boson
internal lines and two Higgs boson internal lines respectively.
For the cc diagram the approximation $\MW \gg m_c$ is still justified.
Thus the results of \Cite{R3,R4} are still valid.
\bigskip

For the ct graph the
terms in Eq. (\ref{eq:1}) can be regrouped in the following way
\begin{eqnarray}
  \label{eq:18}
  H_{WW}^a(c,t) = [ I(0,0) - I(0,m_c^2) ]                \\
  \label{eq:19}
  H_{WW}^b(c,t) = [ I(m_c^2,\mt^2) - I(0,\mt^2) ]
\end{eqnarray}
In Eq. (\ref{eq:18}) which is independent of $m_t$, the W boson can be
 considered again as heavy.  However, since there is only one GIM subtraction,
both W boson propagators  cannot be removed directly.  Instead  the
heavy W limit can be realized  by following the steps leading to
Eq. (\ref{eq:7}), i.~e.

\begin{eqnarray*}
  \int \frac{dp^2}{(p^2+m_c^2)(p^2+\MW^2)^2}  &=&
       \frac1{\MW^2-m_c^2}
       \int\frac{dp^2}{p^2+\MW^2}
           [ \frac1{p^2+m_c^2}  -   \frac1{p^2+\MW^2} ]
     \\ &\approx&
       \frac1{\MW^2}  [
         \frac1{\MW^2}\int_{m_c^2}^{\MW^2} \frac{dp^2}{p^2} -
         \frac1{\MW^2}
         ]
\end{eqnarray*}
where he first intgral recieves contributions from $m_c^2 \le p^2
  \le m_W^2  $ and the second one from $p^2 \approx m_W^2$\cite{R2,R4,R5}.
Thus we obtain the QCD corrected form

\begin{equation}
  \label{eq:20}
  H_{WW}^{a}(c,t) = \frac1{\MW^2}
      \int_{m_c^2}^{\MW^2} \frac{dp^2}{p^2} m_c^2(p^2) Q_1(p^2) Q_2(p^2)
    -m_c^2(\MW^2) Q_2(\MW^2)
\end{equation}

This was already derived in \Cite{R5}, where the number of flavours,
however, was treated in an approximate way as discussed in the last section.
 Following section \ref{secII}(see Eqs. 11 - 17)
 a rigorous result can now be obtained from
Eq. (\ref{eq:20}).The integral is the same as Eq. (\ref{eq:11})
with $\mt$ replaced by $\MW$ and some obvious readjustments  of the
number of flavours.
After these minor modifications the result  agrees completely with the
corresponding formula of \Cite{R6}
which , however, neglects the second term in Eq.( 20) (see below for
further discussions on this point).

For Eq. (\ref{eq:19}), however, the heavy W limit is no longer
justified.  In fact as noted in \Cite{R5},  the corrections in
Eq. (\ref{eq:3}) do not arise any more, because the diagrams with
the gluon connecting an external with an internal quark line decouple
in this case.
Then Eq. (\ref{eq:19}) with QCD corrections becomes

\begin{equation}
  \label{eq:21}
  H_{WW}^b(c,t) \approx - \int \frac{dp^2 m_c^2(p^2) \MW^2}
         {(p^2+\mt^2(p^2))(p^2+\MW^2)^2}
         (\frac{\alpha_{n_2}(p^2)}{\alpha_{n_2}(\mu^2)})  ^ {2/b(n_2)}
\end{equation}

\noindent Here only the leading order term in $m_c^2$ is retained.
For accurate result the integration in Eq.( 21) is done numerically
considering the region $p^2 > \mu^2$ only. In course of the integration
the number of active flavours in the QCD corrections should be adjusted
according to the value of $p^2$ as discussed in section 2.

An analytical formula can be easily obtained   by the  replacement
$\mt^2(p^2) \approx \mt^2$ in Eq.( 21). This corresponds to the approximation
$ (\mt^2(p^2)-\mt^2)/(p^2+\mt^2) \ll 1 $.
Since for the above integral the dominant  contribution comes from
$ \MW^2 \le p^2 \le \mt^2 $,  this approximation is reasonable. Now splitting
the integrand such that different terms receive contributions from
different regions of $p^2$ \cite{R2,R4,R5} we obtain
\begin{eqnarray}
  \label{eq:22}
H_{WW}^b(c,t) &\approx& \int \frac{dp^2 m_c^2 \MW^2}
           {\mt^2-\MW^2}
         [ \frac1{(p^2+\mt^2)(p^2+\MW^2)} - \frac1{(p^2+\MW^2)^2}
         ]  \nonumber\\&&
    (\frac{\alpha_{n_2}(p^2)}{\alpha_{n_2}(\mu^2)})  ^ {2/b(n_2)}
    (\frac{\alpha_{n_3}(p^2)}{\alpha_{n_3}(m_c^2)})  ^ {8/b(n_3)}
               \nonumber\\
            &\approx&     \frac{x_c}{x_t-1}
    (\frac{\alpha_{3}(m_c^2)}{\alpha_{3}(\mu^2)})  ^ {2/b(3)}
    (\frac{\alpha_{4}(m_b^2)}{\alpha_{4}(m_c^2)})  ^ {10/b(4)}
    (\frac{\alpha_{5}(\MW^2)}{\alpha_{5}(m_b^2)})  ^ {10/b(5)}
    \nonumber\\&& 
    [
      \frac1{x_t-1} J(\mt^2,\MW^2,\frac{10}{b_5}) - 1
    ]
\end{eqnarray}
As already discussed (see e.~g.\
Eq. (\ref{eq:9})) the second term in the square bracket of (\ref{eq:22})
can be obtained by evaluating   the QCD corrections at $p^2 \approx \MW^2$.
The first term, however, should be
 evaluated as in Eqs.7, 13 and 14.
This procedure leads to the  approximate result of eq. (\ref{eq:22}).

In \Cite{R6} QCD corrections in the interval $ \MW^2 \to \mt^2 $ were
neglected.    In this approximation ($\alpha(p^2) \approx
\alpha(\MW^2) $ for
$p^2 > \MW^2$), $J$ in Eq. (\ref{eq:22}) reduces to $\ln x_t$ and
the result of \Cite{R6} is regained.  The evolution  above the scale
$\MW$  is given by $J$. However, J multiplies a  term
suppressed by $x_t - 1$ and does
not affect the numerical result appreciably for
$\mt = 174 \GeV$. It should be noted that the new correction is not an
overall multiplicative factor. This feature can be seen in all the
subsequent formulae.

In \Cite{R7} it was suggested   that QCD corrections
to the ct graph can be computed by substituting $\MW \approx \mt$  in
the corresponding formula of \Cite{R3}.  In view of Eq.( 20) and the
discussions following  it, it is clear that this is justified for the part of
$H_{eff}$ given in Eq. (\ref{eq:18}), apart from the neglect of the
second term in Eq.( 20). We have checked that this corresponds to an error of
 6 \%.This prescription is obviously not justified  for the piece in Eq.(19).
 However, the contribution of
Eq. (19) is numerically much smaller than that of Eq. (18).
A posteriori the approximation of \cite{R7} appears to be reasonable.

Following the same procedure
as above one obtains (keeping only the leading order term in $m_c^2$)
for  HH and WH exchanges in the ct diagram:
\begin{eqnarray}
  H_{HH}(c,t) &\approx& \frac{m_c^2 m_t^2}{4\MW^2}
     \int dp^2
     (\frac{\alpha_{n_3}(p^2)}{\alpha_{n_3}(m_c^2)})  ^ {8/b(n_3)}
     (\frac{\alpha_{n_3}(p^2)}{\alpha_{n_3}(\mt^2)})  ^ {8/b(n_3)}
     (\frac{\alpha_{n_2}(p^2)}{\alpha_{n_2}(\mu^2)})  ^ {2/b(n_2)}
     \nonumber\\&& 
       \frac{p^2}{(p^2+\mt^2(p^2))(p^2+\MW^2)^2}
\end{eqnarray}

and
\begin{eqnarray}
H_{WH}(c,t) &\approx& 2 m_c^2 \mt^2 \int dp^2
     (\frac{\alpha_{n_3}(p^2)}{\alpha_{n_3}(m_c^2)})  ^ {8/b(n_3)}
     (\frac{\alpha_{n_3}(p^2)}{\alpha_{n_3}(\mt^2)})  ^ {8/b(n_3)}
     (\frac{\alpha_{n_2}(p^2)}{\alpha_{n_2}(\mu^2)})  ^ {2/b(n_2)}
     \nonumber\\&& 
      \frac1{(p^2+\mt^2(p^2))(p^2+\MW^2)^2}
\end{eqnarray}

\bigskip
The approximation for performing the integrals analytically has
already been described .  Introducing this we arrive at the results
\begin{eqnarray}
  H_{HH}(c,t) &=& \frac{m_c^2 m_t^2}{4\MW^2}
     \int dp^2
     (\frac{\alpha_{n_3}(p^2)}{\alpha_{n_3}(m_c^2)})  ^ {8/b(n_3)}
     (\frac{\alpha_{n_3}(p^2)}{\alpha_{n_3}(\mt^2)})  ^ {8/b(n_3)}
     (\frac{\alpha_{n_2}(p^2)}{\alpha_{n_2}(\mu^2)})  ^ {2/b(n_2)}
     \nonumber\\&& 
     [
        \frac{-\MW^2}{(\mt^2-\MW^2)(p^2+\MW^2)^2}
     +  \frac{\mt^2}{(\mt^2-\MW^2)(p^2+\mt^2)(p^2+\MW^2)}
     ]
     \nonumber\\
       &\approx& \frac{x_c x_t}4
     (\frac{\alpha_{3}(m_c^2)}{\alpha_{3}(\mu^2)})  ^ {2/b(3)}
     (\frac{\alpha_{4}(m_b^2)}{\alpha_{4}(m_c^2)})  ^ {10/b(4)}
     (\frac{\alpha_{5}(\MW^2)}{\alpha_{5}(m_b^2)})  ^ {10/b(5)}
     \nonumber\\&& 
     [
      - \frac1{x_t - 1}
        (\frac{\alpha_{5}(\MW^2)}{\alpha_{5}(\mt^2)})  ^ {8/b(5)}
     \nonumber \\&&\qquad 
      + \frac{x_t}{(x_t-1)^2}
        (\frac{\alpha_{5}(\MW^2)}{\alpha_{5}(\mt^2)})  ^ {8/b(5)}
        J(\mt^2,\MW^2,\frac{18}{b(5)})
        ]\;,
\end{eqnarray}
and
\begin{eqnarray}
H_{WH}(c,t) &=& 2 m_c^2 \mt^2 \int dp^2
     (\frac{\alpha_{n_3}(p^2)}{\alpha_{n_3}(m_c^2)})  ^ {8/b(n_3)}
     (\frac{\alpha_{n_3}(p^2)}{\alpha_{n_3}(\mt^2)})  ^ {8/b(n_3)}
     (\frac{\alpha_{n_2}(p^2)}{\alpha_{n_2}(\mu^2)})  ^ {2/b(n_2)}
     \nonumber\\&& 
     [
       \frac1{(\mt^2-\MW^2)(p^2+\MW^2)^2}
     \nonumber\\&&\qquad 
     -
       \frac1{(\mt^2-\MW^2)(p^2+\mt^2) (p^2+\MW^2)}]
    \nonumber\\
     &\approx&
     \frac{2 x_c x_t}{(x_t-1)}
     (\frac{\alpha_{3}(m_c^2)}{\alpha_{3}(\mu^2)})  ^ {2/b(3)}
     (\frac{\alpha_{4}(m_b^2)}{\alpha_{4}(m_c^2)})  ^ {10/b(4)}
     (\frac{\alpha_{5}(\MW^2)}{\alpha_{5}(m_b^2)})  ^ {10/b(5)}
     \nonumber \\&& 
     [
         (\frac{\alpha_{5}(\MW^2)}{\alpha_{5}(\mt^2)})  ^ {8/b(5)}
      -  \frac1{x_t-1}
         (\frac{\alpha_{5}(\MW^2)}{\alpha_{5}(\mt^2)})  ^ {8/b(5)}
         J(\mt^2,\MW^2,\frac{18}{b(5)})
     ] \nonumber\\
\end{eqnarray}

\bigskip
\noindent As has already been stated,
they give good estimates, but for very accurate  computation it is
desirable to integrate the exact equations numerically. The QCD
corrections from $\mt \to \MW$  are contained in the square brackets.
The three overall factors give the QCD evolution from
          $\mu\to\MW$, and agrees with \Cite{R6,R8}.

\medskip
Finally,  we consider the tt graphs.  For reasons discussed earlier,
the correction in Eq. (\ref{eq:3}) are no longer relevant.
Including the other two types of corrections we obtain  for the  WW,
HH  and WH  exchanges

\begin{eqnarray}
  H_{WW}(t,t) &=&   \MW^2 \int dp^2 \mt^4( p^2 )
      (\frac{\alpha_{n_2}(p^2)}{\alpha_{n_2}(\mu^2)})  ^ {2/b(n_2)}
     \nonumber\\&& 
          \frac1{(p^2+\mt^2(p^2))^2 (p^2+\MW)^2}
\end{eqnarray}

\medskip
\begin{eqnarray}
  H_{HH}(t,t) &=& \frac{\mt^4}{4\MW^2}
      \int dp^2
      (\frac{\alpha_{n_2}(p^2)}{\alpha_{n_2}(\mu^2)})  ^ {2/b(n_2)}
      (\frac{\alpha_{n_3}(p^2)}{\alpha_{n_3}(\mt^2)})  ^ {16/b(n_3)}
     \nonumber\\&& 
          \frac{p^4}{(p^2+\mt^2(p^2))^2 (p^2+\MW)^2}
\end{eqnarray}

\medskip
\begin{eqnarray}
  H_{WH}(t,t) &=& 2\mt^4
      \int dp^2
      (\frac{\alpha_{n_2}(p^2)}{\alpha_{n_2}(\mu^2)})  ^ {2/b(n_2)}
      (\frac{\alpha_{n_3}(p^2)}{\alpha_{n_3}(\mt^2)})  ^ {16/b(n_3)}
     \nonumber\\&& 
         \frac{p^2}{(p^2+\mt^2(p^2))^2 (p^2+\MW)^2}
\end{eqnarray}

\medskip
\noindent
The two types of QCD corrections arise from the running top quark mass
and the diagrams in fig. 2.  They produce the terms containing
$b(n_3)$ and $b(n_2)$, respectively.  Again the numbers of flavors
should be assigned consistently at each momentum interval.  Thus
these formulae with the appropriate choice of the number of generations
$n_2,n_3$ and $\mu$ can be used for computing the contributions of the
box diagrams for K- and B-mesons.

Introducing again the approximation described after Eq. (21) we
arrive at the formulae :

\begin{eqnarray}
  H_{WW}(t,t) &=&  \mt^4 \MW^2 \int \frac{dp^2}{(\mt^2 - \MW^2)^2}
      (\frac{\alpha_{n_2}(p^2)}{\alpha_{n_2}(\mu^2)})  ^ {2/b(n_2)}
     \nonumber\\&& 
      [
          \frac1{(p^2+\mt^2)^2}
        + \frac1{(p^2+\MW)^2}
        - 2 \frac1{(p^2+\MW^2)(p^2+\mt^2)}
      ]\;,\nonumber\\ 
      & \approx &
      (\frac{\alpha_{3}(m_c^2)}{\alpha_{3}(\mu^2)})  ^ {2/b(3)}
      (\frac{\alpha_{4}(m_b^2)}{\alpha_{4}(m_c^2)})  ^ {2/b(4)}
      (\frac{\alpha_{5}(\MW^2)}{\alpha_{5}(m_b^2)})  ^ {2/b(5)}
     \nonumber\\&& 
      [
         \frac{x_t}{(x_t-1)^2}
          (\frac{\alpha_{5}(\mt^2)}{\alpha_{5}(\MW^2)})  ^ {2/b(5)}
        +
        \frac{x_t^2}{(x_t-1)^2}
     \nonumber\\&&\qquad 
        -2
        \frac{x_t^2}{(x_t-1)^3}
          J(\mt^2,\MW^2,\frac{2}{b_5})
      ]\;,
\end{eqnarray}
Special care should be taken in deriving eq.\ (30) from eq.\ (27).
The $m_t^4(p^2)$ in Eq. (27) originates from the top propagator due to GIM
subtraction.
Consequently, when we neglect the running of the top quark mass in the
propagator, we must consistently neglect the running of the
$m^4_t(p^2)$-terms.
Keeping this in mind we also
obtain

\medskip
\begin{eqnarray}
  H_{HH}(t,t) &=& \frac{\mt^4}{4\MW^2}
      \int \frac{dp^2}{(\mt^2 - \MW^2)^2}
      (\frac{\alpha_{n_2}(p^2)}{\alpha_{n_2}(\mu^2)})  ^ {2/b(n_2)}
      (\frac{\alpha_{n_3}(p^2)}{\alpha_{n_3}(\mt^2)})  ^ {16/b(n_3)}
     \nonumber\\&& 
      [
          \frac{\mt^4}{(p^2+\mt^2)^2}
        + \frac{\MW^4}{(p^2+\MW)^2}
        - 2 \frac{\MW^2\mt^2}{(p^2+\MW^2)(p^2+\mt^2)}
      ]\;,\nonumber\\ 
      & \approx &
      (\frac{\alpha_{3}(m_c^2)}{\alpha_{3}(\mu^2)})  ^ {2/b(3)}
      (\frac{\alpha_{4}(m_b^2)}{\alpha_{4}(m_c^2)})  ^ {2/b(4)}
      (\frac{\alpha_{5}(\MW^2)}{\alpha_{5}(m_b^2)})  ^ {2/b(5)}
     \nonumber\\&& 
      [
         \frac{x_t^3}{4(x_t-1)^2}
          (\frac{\alpha_{5}(\mt^2)}{\alpha_{5}(\MW^2)})  ^ {2/b(5)}
        +
        \frac{x_t^2}{4(x_t-1)^2}
          (\frac{\alpha_{5}(\MW^2)}{\alpha_{5}(\mt^2)})  ^ {16/b(5)}
        -
     \nonumber\\&& 
        \frac{x_t^3}{2(x_t-1)^3}
          (\frac{\alpha_{5}(\MW^2)}{\alpha_{5}(\mt^2)})  ^ {16/b(5)}
          J(\mt^2,\MW^2,\frac{18}{b_5})
      ]\;,
\end{eqnarray}

\medskip
\begin{eqnarray}
  H_{WH}(t,t) &=& 2\mt^4
      \int \frac{dp^2}{(\mt^2 - \MW^2)^2}
      (\frac{\alpha_{n_2}(p^2)}{\alpha_{n_2}(\mu^2)})  ^ {2/b(n_2)}
      (\frac{\alpha_{n_3}(p^2)}{\alpha_{n_3}(\mt^2)})  ^ {16/b(n_3)}
     \nonumber\\&& 
      [
        - \frac{\mt^2}{(p^2+\mt^2)^2}
        - \frac{\MW^2}{(p^2+\MW)^2}
     \nonumber\\&&\qquad 
        + \frac{\mt^2 + \MW^2}{(p^2+\MW^2)(p^2+\mt^2)}
      ]\;     \nonumber\\ 
       & \approx &
      (\frac{\alpha_{3}(m_c^2)}{\alpha_{3}(\mu^2)})  ^ {2/b(3)}
      (\frac{\alpha_{4}(m_b^2)}{\alpha_{4}(m_c^2)})  ^ {2/b(4)}
      (\frac{\alpha_{5}(\MW^2)}{\alpha_{5}(m_b^2)})  ^ {2/b(5)}
      \frac{2x_t^2}{(x_t-1)^2}
     \nonumber\\&& 
      [
         \frac{(1 + x_t)}{(x_t-1)}
            (\frac{\alpha_{5}(\MW^2)}{\alpha_{5}(\mt^2)})  ^ {16/b(5)}
            J(\mt^2,\MW^2,\frac{18}{b(5)})
       \nonumber\\&&
        - (\frac{\alpha_{5}(\MW^2)}{\alpha_{5}(\mt^2)})  ^ {-2/b(5)}
        -
          (\frac{\alpha_{5}(\MW^2)}{\alpha_{5}(\mt^2)})  ^ {16/b(5)}
      ]\;.
\end{eqnarray}

\bigskip
\noindent
We note that the three overall factors in Eqs.\ (30) - (32) give the QCD
evolution from $\MW \rightarrow \mu$ as was already noted in \Cite{R6,R8}.
The results of \Cite{R6} are obtained by setting $\alpha(\mt^2)
= \alpha(\MW^2)$ and $J = ln \; x_t$ in Eq. (30) - (32).

For $B-\bar{B}$ mixing only tt diagrams are relevant. QCD corrections
can be obtained by carrying out the $p^2$ integrations in Eqs.
 (27)- (29) numerically. However, in this case $\mu = 0(m_B)$. Hence only $p^2
> m_b^2$ should be considered. The analytical formulae in this  case are
\begin{eqnarray}
  H_{WW}(t,t) &\approx&
      (\frac{\alpha_{5}(\MW^2)}{\alpha_{5}(\mu^2)})  ^ {2/b(5)}
      \frac{x_t}{(x_t-1)^2}
      [
          (\frac{\alpha_{5}(\mt^2)}{\alpha_{5}(\MW^2)})  ^ {2/b(5)}
        +  x_t
        \nonumber\\&&
        -2
        \frac{x_t}{(x_t-1)}
          J(\mt^2,\MW^2,\frac{2}{b_5})
      ]\;,
\end{eqnarray}

\medskip
\begin{eqnarray}
  H_{HH}(t,t)
      & \approx &
      (\frac{\alpha_{5}(\MW^2)}{\alpha_{5}(\mu^2)})  ^ {2/b(5)}
      \frac{x_t^2}{4(x_t-1)^2}
      [
         x_t (\frac{\alpha_{5}(\mt^2)}{\alpha_{5}(\MW^2)})  ^ {2/b(5)}
        +
          (\frac{\alpha_{5}(\MW^2)}{\alpha_{5}(\mt^2)})  ^ {16/b(5)}
       \nonumber\\&&
        -
        \frac{2 x_t}{(x_t-1)}
          (\frac{\alpha_{5}(\MW^2)}{\alpha_{5}(\mt^2)})  ^ {16/b(5)}
          J(\mt^2,\MW^2,\frac{18}{b_5})
      ]\;,
\end{eqnarray}

\medskip
\begin{eqnarray}
  H_{WH}(t,t)
       & \approx &
      (\frac{\alpha_{5}(\MW^2)}{\alpha_{5}(\mu^2)})  ^ {2/b(5)}
      \frac{2x_t^2}{(x_t-1)^2}
     \nonumber\\&& 
      [
         \frac{(1 + x_t)}{(x_t-1)}
            (\frac{\alpha_{5}(\MW^2)}{\alpha_{5}(\mt^2)})  ^ {16/b(5)}
            J(\mt^2,\MW^2,\frac{18}{b(5)})
       \nonumber\\&&
        - (\frac{\alpha_{5}(\MW^2)}{\alpha_{5}(\mt^2)})  ^ {-2/b(5)}
        -
          (\frac{\alpha_{5}(\MW^2)}{\alpha_{5}(\mt^2)})  ^ {16/b(5)}
      ]\;.
\end{eqnarray}
We are now prepared to discuss the numerical results.

\section{Numerical Results and Conclusions}
\noindent
The main results of this article are QCD corrected effective Hamiltonians
in the LLA for $K^0 - \bar{K^0}$ and  $B^0 - \bar{B^0}$ systems  given by
the formulae (23)--(35) in the case $m_t > m_W$. We present for the first
time analytical formulae which show explicitly the effect of QCD evolutions
in the interval $m_W \longrightarrow m_t$, neglected by
other authors. Further we have shown the agreement between different
calculations which apparently use different methods.

All parameters in the formulae are defined except for the infrared
scale $\mu$, which is a substraction point.  For physical states
the low energy scale must be set by the masses of the quarks and
the dynamics of confinement.  A reasonable expectation is $\mu =0(m_K)$
for K-mesons and $\mu = 0(m_B)$ for B-mesons.  This, of course, means
that the renormalization for B-mesons is stopped at $M_B$, while for
K-mesons is continued down to and below the $m_c$ threshold.  For this reason
formulae for the two cases are different:  for B-mesons we do not
cross the $m_c$ threshold.

The $\mu$-dependence brings in a factor $(\alpha_{n_2} (\mu^2))^{-2/b(n_2)}$
which should be cancelled, in principle, by a corresponding factor in the
hadronic matrix element.  This motivated some authors \cite{R8,R9} to
factor out such a term and define the QCD factors without it.  This does
not solve the problem but pushes it to another part of the theory; up to
now there is no explicit calculation where the $\mu$-dependence of the
matrix element cancels the $\mu$-dependence from QCD.  For the K-meson
we will take $\alpha_s(m^2_K) \approx 0.29$
and then take the 6/25 root of it,
which brings the $\mu$-term closer to 1.  For B-mesons we will take
$\mu=0(m_B)$.

We use the formulae of section 3 to calculate the integrals for various
values of a heavy top quark.
We present our results in several tables.  Table 1 shows the
corrections for the tt-graphs of $K^0 - \bar{K^0}$ mixing
for $\mt = 174 \GeV, m_c = 1.3 \GeV,
m_b = 4.49 \GeV, \MW = 80 \GeV$ and $\mu = 0.50 \GeV $. Throughout the
paper the $\Lambda parameter$ of QCD with three active flavours will be
taken as input with a value $\Lambda_3 =0.3$ which is consistent with
the value of $\Lambda_5$ in \cite{R12}. This parameter for other
numbers of active flavours  will be determined fron the input value
via the matching conditions \cite{R3}.
We show three
terms of the Hamiltonian with WW, WH and HH exchanges, respectively.
The first column is the box diagram with only weak terms.  The second
column gives the values when QCD corrections are included and
the integrals in Eqs (27) - (29) are evaluated numerically.  As we discussed
in the previous section, the evolution of $\alpha_s( p^2 )$ for
$p^2 > \MW^2$ is now included.
We notice that the change from the first to the second column is large,
giving an
\begin{displaymath}
\eta_2=0.56.
\end{displaymath}
This means that, as already  noted by previous  works, the strong
contributions are sizeable and must be included.

In the previous sections, we also demonstrated that we can calculate
the integrals analytically by introducing an approximation.  We
calculated these terms and found out that they are in reasonable agreement
with the numbers obtained by numerical integration. The analytical
formulae are particularly useful for studying the evolution above
$\MW$ because the effects of this evolution   is explicitly
separated.  The separation shows that the evolution above $\MW$ is not
quite negligible  when corrections to individual terms in Eqs. (30) - (32)
are considered. For the purpose of illustration we show below the numerical
values of the QCD corrections in Eq. (31).
\begin{eqnarray}
  H_{HH}(t,t)
      & \approx & 0.58 \frac{x_t^2}{4(x_t-1)^2}
     \nonumber\\&& 
      [
       0.97  x_t + 1.29 -  1.12  \frac{2 x_t}{x_t-1} \ln x_t
      ]\;,
\end{eqnarray}
In  Eq. (36)  the pure electroweak term is kept unmodified, i.e. setting
all numerical factors equal to one we recover the electroweak term. The
factor 0.58 is due to the QCD correction in the interval $\mu^2
\longrightarrow m_W^2$ \cite{R6,R7,R8}.
The effects of the evolution in the interval $m_W \longrightarrow m_t$
are shown explicitly as a modification of the coefficients in the square
bracket. The  bulk of these effects
comes from the running of the top mass which in this case
 has significantly  larger anomalous dimension compared
to the corrections in Eq. (4). The top quark mass correction is
absent in the first term in Eq. (36) since this term is dominated by
$p^2 \approx m_t^2$ (see the first step of Eq. (31)).
The second  term which is dominated by
$p^2 \approx m_W^2$ (see the first step of Eq. (31)) receives the maximum
correction (1.29) , as expected. But the  overall numerical impact of this
term is rather modest since it multiplies a  small electroweak term.
The complete results are shown in table 2. The first column gives the
results from the full analytical formulae( Eqs. (30) - (32)) while the
second column is obtained by ignoring the evolutions above $m_W$.
We also  performed  the numerical integration  in the latter approximation
(i.e, by setting $\alpha(p^2) = \alpha(\MW^2)$ for $\MW^2 < p^2$).  The
results are given in the third column of table 1 and are in fair agreement with
the analytical results.
  The corrections for the
$H_{WW}$ and $H_{HH}$ terms differ from those on the second column,
but the deviations are in opposite directions so that they partly
cancel in the sum.

In Table 3 we show the results for the ct-diagrams for $m_t=174\GeV$.
We note that the largest contribution comes from the $H^a_{WW}$ term (Eq.
 (20))whose integral is dominated by the region
$m_c^2 \le p^2   \le m_W^2  $ .  All
other terms are smaller. We obtain the correction  $\eta_3 $ in agreement
with previous results.

Table 4 shows the corrections for the tt-graphs of the K-mesons for
$\mt = 150$ and $200 \GeV/c^2$. We show again three terms of the Hamiltonian
with WW, WH and HH exchanges.  We emphasize that the values for $\eta_2$
are the same demonstrating, a posteriori, that a precise definiton of
the running $m_t$ mass is not crucial as long as it is with in the range
permitted by the experiments \cite{R10,R11}.

We did not calculate the ct-graphs for other values of $\mt$ because
the dominant term by far is $H^a_{WW}$ which is independent of $\mt$.
In summary we see that the QCD corrections to box diagrams of the
K-system are well
understood and give precise values.  The values for the $K^0 - \bar{K}^0$
system are
\begin{displaymath}
\eta_2 = 0.56 \qquad \rm{and} \qquad \eta_3 = 0.36 \;.
\end{displaymath}

We computed the effective Hamiltonians for the B-system also for three
masses of the top quark $m_t = 150, 174$ and $200 \GeV$ and for the
parameters $m_c = 1.3 GeV$, $m_b = 4.49 GeV$, $\MW = 80 \GeV$ and
$\mu = 4.49 \GeV$.  In Table 5 we show the three terms of the Hamiltonian
for $WW, WH$ and $HH$ exchanges, separately.  We note that $\eta_2$ is
now larger than the K system.
\begin{displaymath}
\eta_2 = 0.81
\end{displaymath}
and stable on various mass of $\mt$.  A large part of the change comes
from the change of the scale $\mu$.In the last column of Table 5 we
give the results for a hypothetical T quark belonging to the fourth
generation. In this case $\eta_2$ changes perceptibly. Ignoring the
evolution in the interval $m_W \longrightarrow m_T$ one obtains $\eta_2
= 0.84$.  We have also verified that the analytical formulae
Eqs. (33) - (35) give numbers in agreement with Table 5.

We can compare these results with other calculations.
\begin{enumerate}
\item \underline{For the K-system}. The value of $\eta_2$ is practically
the same as in our earlier paper \cite{R5}.  For $\eta_2$ there is a
next-to-leading order calculation \cite{R8} which gives again the same
value.  We have shown in this article that $\eta_2$ is stable under
variations of the top quark mass between 150 and 200 \GeV.  For $\eta_3$
the new value is in agreement with  our earlier result \cite{R5}.
\item  \underline{For the B-system} the dominant contribution comes from
the tt-diagrams.  The $\eta_2$ is larger than in the $K^0$-mesons.  This
comes from the change in the $\mu$-scale and the higher threshold.
In comparing with \Cite{R8} one must be careful to correct for the
$(\alpha_{\eta_2}(\mu^2))^{2/b(\eta_2)}$ which is included in our formulae.
In \Cite{R8} this factor is extracted and consequently the value for
$\eta_{2B}$ is practically equal to $\eta_{2K}$.
\end{enumerate}

In conclusion, the QCD factors which enter calculations with box diagrams
have stabilized over the past few years.  Improvements through the
inclusion of thresholds and variations on the mass of the top quarks
discussed in these papers change the values very little or not at all.
The QCD factors are much more stable and better understood than the other
parameters, in particular the Cabibbo-Kobayashi-Maskawa matrix elements
and reduced matrix elements, which enter these calculations.

\bigskip
\bf{Acknowledgements}: The work of AD was supported partly by the
Alexander von Humboldt Stiftung, Germany and partly by the Department
of Science and Technology, Government of India.  We wish to thank the
Bundesministerium fuer Bildung, Wissenschaft, Forschung und Technologie
(BMBF), Bonn, for financial support under Contract No. 056DO93P(5).

\clearpage

\newpage
\begin{table}
\begin{center}
\begin{tabular}{||l|c|c|c||}
\hline \hline
& without & with QCD & with QCD \\
&& numerically & negl. above $\MW$ \\
\hline
$H_{WW}$ & 0.61 & 0.37 & 0.36 \\
$H_{WH}$ & 1.25 & 0.72 & 0.73 \\
$H_{HH}$ & 0.72 & 0.36 & 0.42 \\
\hline
sum & 2.58 & 1.45 & 1.51 \\
$\eta_2=$ & & 0.56 & 0.59 \\
\hline \hline
\end{tabular}
\end{center}
\caption{QCD corrections for the tt-diagrams for $\mt = 174 \GeV/c^2$}
\end{table}

\vspace{5.0cm}
\begin{table}
\begin{center}
\begin{tabular}{||l|c|c||}
\hline \hline
&  with QCD(A) & with QCD(B) \\
\hline
$H_{WW}$ & 0.36 & 0.35 \\
$H_{WH}$ & 0.77 & 0.72 \\
$H_{HH}$ & 0.34 & 0.42 \\
\hline
sum & 1.47 & 1.49 \\
$\eta_2=$ & 0.57 & 0.58 \\
\hline \hline
\end{tabular}
\end{center}
\caption{Corrections for the tt-diagrams for $\mt = 174 \GeV/c^2.$
A) Results from full analytical formulae. B) Results after neglecting
the evolution above $m_W$.}
\end{table}

\vspace{5.0cm}
\begin{table}
\begin{center}
\begin{tabular}{||l|c|c|c||}
\hline \hline
& without & with QCD & with QCD \\
&& numerically & negl. above $\MW$ \\
\hline
$H^a_{WW}$ & 1.91 & 0.78 & 0.78 \\
$H^b_{WW}$ & -0.041 & -0.006 & -0.007 \\
$H_{WH}$ & 0.39 & 0.072 & 0.074 \\
$H_{HH}$ & 0.081 & 0.012 & 0.014 \\
\hline
sum & 2.34 & 0.86 & 0.86 \\
$\eta_3=$ & & 0.37 & 0.37 \\
\hline \hline
\end{tabular}
\end{center}
\caption{Corrections for the ct-diagrams in units $10^{-3}$ for
			$\mt = 174 \GeV/c^2$}
\end{table}

\newpage

\begin{table}
\begin{center}
\begin{tabular}{||l|c|c|c|c||}
\hline \hline
& \multicolumn{2}{|c|}{For $\mt = 150 \GeV/c^2$}
   & \multicolumn{2}{|c||}{For $\mt = 200 \GeV/c^2$} \\
\hline
& without & with QCD & without & with QCD \\
&& numerically && numerically \\
\hline
$H_{WW}$ & 0.56 & 0.34 & 0.65 & 0.40\\
$H_{WH}$ & 1.00 & 0.58 & 1.50 & 0.87 \\
$H_{WW}$ & 0.49 & 0.24 & 1.02 & 0.51 \\
\hline
sum & 2.05 & 1.16 & 3.17 & 1.78 \\
$\eta_2=$ & \multicolumn{2}{|c|}{0.57}
   & \multicolumn{2}{|c||}{0.56} \\
\hline \hline
\end{tabular}
\end{center}
\caption{The tt-case in units of 1 for values of $\mt$}
\end{table}

\vspace{5.0cm}
\begin{table}
\begin{center}
\begin{tabular}{||l|c|c|c|c|c|c|c|c||}
\hline \hline
& \multicolumn{2}{|c|}{$\mt = 150 \GeV/c^2$}
   & \multicolumn{2}{|c|}{$\mt = 174 \GeV/c^2$}
      & \multicolumn{2}{|c||}{$\mt = 200 \GeV/c^2$}
         & \multicolumn{2}{|c|}{$\mt = 500 \GeV/c^2$}\\
\hline
& without & with QCD & without & with QCD &  without & with QCD
&  without & with QCD \\
\hline
$H_{WW}$ & 0.56 & 0.49 & 0.61 & 0.53 & 0.65 & 0.57& 0.87 & 0.75\\
$H_{WH}$ & 1.00 & 0.84 & 1.25 & 1.04 & 1.50 & 1.26& 3.91 & 3.23 \\
$H_{HH}$ & 0.49 & 0.35 & 0.72 & 0.52 & 1.02 & 0.74& 8.55 &6.29 \\
\hline
sum & & & & & & \\
$\eta_2$ & \multicolumn{2}{|c|}{0.82}
            & \multicolumn{2}{|c|}{0.81}
               & \multicolumn{2}{|c||}{0.81}
                  & \multicolumn{2}{|c||}{0.77}\\
\hline \hline
\end{tabular}
\end{center}
\caption{The tt-case for B-mesons in units of 1}
\end{table}
\end{document}